\documentstyle[preprint,aps,here,psfig]{revtex}
\begin{document}
\tightenlines
\preprint{HUPD-9722}
\title{Effect of $\rho N$ channel in the $\gamma N\rightarrow \pi \pi N$
reactions}
\author{
M. Hirata\thanks{E-mail:
hirata@theo.phys.sci.hiroshima-u.ac.jp},
K. Ochi\thanks{E-mail:
kazu@theo.phys.sci.hiroshima-u.ac.jp}
\\
{\it Department of Physics, Hiroshima University, Higashi-Hiroshima
739, Japan}
\\
and
\\
T. Takaki\thanks{E-mail:
takaki@hiroshima-cdas.or.jp}
\\
{\it Onomichi Junior College, Onomichi 722, Japan}}
\maketitle
\begin{abstract}
 A model for the two-pion photoproduction on the nucleon proposed
earlier
is modified to simultaneously explain the total cross sections and the
invariant mass spectra. Using this model, we discuss the role of the
$\rho $
meson in the $\gamma N\rightarrow \pi \pi N$ reaction. 
\\
\\
PACS number(s): 25.20.Lj, 25.20.Dc, 13.60.Rj\\
\end{abstract}
\newpage
Recently, the two-pion photoproduction on the nucleon have been
experimentally studied for the photon energy from 450 to 800 MeV at
Mainz
Microtron MAMI\cite{exp1,exp2,exp3}. The total cross sections of 
the $\gamma p\rightarrow \pi^{+}\pi ^{0}n$ and $\gamma p\rightarrow 
\pi ^{0}\pi ^{0}p$ reactions have
been
obtained for the first time using the large acceptance detector DAPHNE
and
high intensity tagged photon beams\cite{exp1,exp2}. 
The $\gamma p\rightarrow \pi ^{+}\pi ^{-}p
$ and $\gamma n\rightarrow \pi ^{-}\pi ^{0}p$ cross sections have been
also
measured with good accuracy. Then, $\gamma p\rightarrow \pi
^{0}\pi ^{0}p$ cross sections have been measured using the Glasgow
Tagger
and the TAPS photon spectrometer\cite{exp3}and the previous
 experimental result has
been confirmed. A characteristic feature in this energy region is that
the
resonances such as $\Delta (1232)$ and $N^{*}(1520)$ are involved in the
production process.

DAPHNE-experiments have motivated several
authors\cite{oset,laget,ochi}to
develop the model for the $\gamma N\rightarrow \pi \pi N$ reaction. The
theoretical studies for the $\gamma p\rightarrow \pi ^{+}\pi ^{-}p$
reaction
have shown that the two-pion photoproduction takes place dominantly 
through the 
$\pi \Delta (1232)$ intermediate state, which arises from the $\Delta $
Kroll-Ruderman and $\Delta $ pion-pole terms [Figs. 1(a)-1(b)] and
the $N^{*}(1520)$ excitation [Fig. 1(c)]. The interference between
the $\Delta $ Kroll-Ruderman and the $N^{*}(1520)$ excitation processes
is essential to reproduce the energy dependence of the total cross
section\cite{oset}.

However, it has been found that the neutral pion production such as the
$\gamma
p\rightarrow \pi ^{+}\pi ^{0}n$ and $\gamma n\rightarrow \pi ^{-}\pi
^{0}p$
cannot be explained with only the $\pi \Delta (1232)$ production
mechanism which dominates the $\gamma p\rightarrow \pi ^{+}\pi ^{-}p$
reaction and therefore some additional mechanism is needed. In fact, the
magnitude of cross sections is largely underestimated compared with the
data\cite{oset,laget}. In our previous paper\cite{ochi}, 
we have proposed a simple model
which is able to explain the data and indicated 
that the $\rho N$ intermediate 
state arising from both the $N^{*}(1520)$ 
excitation [Fig. 1(d)] and the $\rho$ Kroll-Ruderman process 
[Fig. 1(e)] plays an important role in the $\gamma p\rightarrow
\pi ^{+}\pi ^{0}n$ and $\gamma n\rightarrow \pi ^{-}\pi ^{0}p$
reactions. 

We note
that the transitions to the $\pi \Delta$ channel in these reactions, 
especially the $\Delta$ Kroll-Ruderman and $\Delta$ pion-pole processes,
are suppressed compared with the $\gamma p\rightarrow \pi ^{+}\pi ^{-}p$ 
reaction because of the isospin factors.

 In addition to the total cross sections of the two-pion
photoproduction, 
the measurements of the invariant mass spectra on the $\gamma
n\rightarrow
\pi ^{-}\pi ^{0}p$ reaction have been performed at Mainz
lately\cite{exp4}.
This experimental result can provide an additional constraint on the
theoretical model. In this letter, we report a modified version of our
model
and discuss our results concerning the total cross sections and
the
invariant mass spectra.

First of all, we review the formalism of our model briefly\cite{ochi}.
The $T$ matrix
for the two-pion photoproduction is written as 
\begin{eqnarray}
T=T_{\Delta KR}+T_{\Delta PP}+T_{N^{*}\pi \Delta }^{s}+T_{N^{*}\pi
\Delta
}^{d}+T_{N^{*}\rho N}+T_{\rho KR}. 
\end{eqnarray}
The $T$ matrix includes two dominant channels, i.e., the $\pi \Delta
(1232)$
and $\rho N$ channels. These states are assumed to arise from six
processes described by the $\Delta $ Kroll-Ruderman term ($T_{\Delta
KR}$%
), $\Delta $ pion-pole term ($T_{\Delta PP}$), $N^{*}(1520)$ excitation
terms ($T_{N^{*}\pi \Delta }^{s(d)}$ and $T_{N^{*}\rho N}$), and
$\rho$ Kroll-Ruderman term ($T_{\rho KR}$) which are shown in Figs. 1
(a)-(e), respectively. The $N^{*}(1520)$ decay into a $\pi \pi N$
occurs through three channels: the $s$-wave $\pi \Delta (1232)$,
$d$-wave $\pi \Delta (1232)$ and $\rho N$ channels. 
The branching fractions into these
decay channels are comparable.

The $\Delta $ Kroll-Ruderman and $\Delta $ pion-pole terms are written
as

\begin{eqnarray}
T_{\Delta KR} &=&F_{\pi N\Delta }G_{\pi \Delta }(s,\vec{p}_{\Delta
})F_{\Delta KR}^{\dagger }, \\
T_{\Delta PP} &=&F_{\pi N\Delta }G_{\pi \Delta }(s,\vec{p}_{\Delta
})F_{\Delta PP}^{\dagger },
\end{eqnarray}
where 
\begin{eqnarray}
G_{\pi \Delta }(s,\vec{p}_{\Delta })=\frac{1}{\sqrt{s}-\omega _{\pi
}(\vec{p}%
_{\Delta })-E_{\Delta }(\vec{p}_{\Delta })-\Sigma _{\Delta }^{(\pi
N)}(s,%
\vec{p}_{\Delta })}. 
\end{eqnarray}
Here, $F_{\pi N\Delta }$ is the $\pi N\Delta $ vertex function which is
taken to be the same vertex function used in the Betz-Lee
model\cite{betz}. 
$G_{\pi
\Delta }(s,\vec{p}_{\Delta })$ is the propagator of the $\pi \Delta $
system, $\Sigma _{\Delta }^{(\pi N)}(s,\vec{p}_{\Delta })$ is the $%
\Delta $ self-energy with the momentum $\vec{p}_{\Delta }$, and $\omega
_{\pi }(\vec{p}_{\Delta })$ and $E_{\Delta }(\vec{p}_{\Delta })$ are the
energies of pion and $\Delta $ , respectively. $F_{\Delta KR}^{\dagger
}$ is
the $\Delta $ Kroll-Ruderman vertex which are obtained from the $%
N\rightarrow \pi \Delta $ vertex function by requiring gauge invariance.
This $N\rightarrow \pi \Delta $ vertex function is assumed to be the
same
form with the $\Delta \rightarrow \pi N$ vertex function of the Betz-Lee
model. The range parameter of the form factor $Q_{\Delta }(N\rightarrow
\pi
\Delta )$ is, however, varied and determined to fit the $\gamma
p\rightarrow
\pi
^{+}\pi ^{-}p$ cross section. $F_{\Delta PP}^{\dagger }$ is 
the $\Delta $ pion-pole vertex.
The $N^{*}(1520)$ terms are written as 
\begin{eqnarray}
T_{N^{*}\pi \Delta }^{s(d)} &=&F_{\pi N\Delta }G_{\pi \Delta
}(s,\vec{p}%
_{\Delta })F_{\pi \Delta N^{*}}^{s(d)}G_{N^{*}}(s)\tilde{F}_{\gamma
NN^{*}}^{\dagger }, \\
T_{N^{*}\rho N} &=&F_{\rho \pi \pi }G_{\rho N}(s,\vec{q}_{\rho })F_{\rho
NN^{*}}G_{N^{*}}(s)\tilde{F}_{\gamma NN^{*}}^{\dagger },
\end{eqnarray}
where 
\begin{eqnarray}
G_{N^{*}}(s) &=&\frac{1}{\sqrt{s}-M_{N^{*}}-\Sigma _{N^{*}}(s)}, \\
G_{\rho N}(s,\vec{q}_{\rho }) &=&\frac{1}{2\omega _{\rho }(\vec{q}_{\rho
})[%
\sqrt{s}-\omega _{\rho }(\vec{q}_{\rho })-E_{N}(\vec{q}_{\rho })-\Sigma
_{\rho \pi \pi }(s,\vec{q}_{\rho })]}.
\end{eqnarray}
Here, $\tilde{F}_{\gamma NN^{*}}^{\dagger }$ is the $\gamma NN^{*}$
vertex
function. $F_{\pi \Delta
N^{*}}^{s(d)}$ is the $\pi \Delta N^{*}$ vertex function for the $s$- or
$d$-wave $\pi \Delta$ state
and $F_{\rho NN^{*}}$ is the $\rho NN^{*}$ vertex function,
respectively.
$F_{\rho \pi \pi }$ is the $\rho \pi \pi $ vertex function.
$G_{N^{*}}(s)$ and $G_{\rho
N}(s,\vec{q}_{\rho })$ are the propagators of the $N^{*}(1520)$ and 
$\rho N$
system, respectively. $\Sigma _{N^{*}}(s)$ is the $N^{*}$ self-energy in
the
center of mass system and $\Sigma _{\rho \pi \pi }(s,\vec{q}_{\rho })$ 
is the $\rho $ meson
self-energy with the momentum $\vec{q}_{\rho }$. $M_{N^{*}}$ is the bare
mass of $N^{*}$ and $\omega _{\rho }(\vec{q}_{\rho })$ and
$E_{N}(\vec{q}%
_{\rho })$ are the energies of the $\rho $ meson and nucleon,
respectively.
The $\rho$ Kroll-Ruderman term is written as 
\begin{eqnarray}
T_{\rho KR}=F_{\rho \pi \pi }G_{\rho N}(s,\vec{q}_{\rho })F_{\rho KR}
^{\dagger }, 
\end{eqnarray}
where $F_{\rho KR}$ is the $\rho$ Kroll-Ruderman vertex which is
derived from the non-relativistic $\rho NN$ vertex function by requiring
gauge
invariance.\allowbreak The detailed forms of the above vertex functions
are
given in Ref.\cite{ochi}. 
The self-energies of the $N^{*}$, $\pi \Delta $ system and $\rho
N$ 
system in the propagators are obtained by using these strong vertex
functions whose expressions are also given in Ref.\cite{ochi}.

Most of the parameters such as coupling constants, range parameters and
bare
masses are phenomenologically obtained by using the $\pi N$ scattering
amplitudes, $\gamma N\rightarrow \pi N$ multipole amplitudes, branching
fractions of the $N^{*}(1520)$ and width of the $\rho $-meson, 
but the signs of the coupling constants and the range parameters 
such as $Q_{\Delta }(N\rightarrow \pi \Delta )$ and $q_{\rho \pi \pi }$
are
not determined. 
Here $q_{\rho \pi \pi }$ is the
range parameter of the $\rho \pi \pi$ form factor. 
The $\gamma N\rightarrow \pi \pi
N $ reaction data is necessary to fix the signs and these remaining
parameters. In order to fix these parameters, we took the following way:
The
signs of the coupling constants and the range parameter $Q_{\Delta
}(N\rightarrow \pi \Delta )$ were determined so as to reproduce the
$\gamma
p\rightarrow \pi ^{+}\pi ^{-}p$ cross section. The range parameter
$q_{\rho
\pi \pi }$ was varied to fit the $\gamma
p\rightarrow \pi ^{+}\pi ^{0}n$ cross section, since the $\rho N$
channel
contributes to the $\gamma p\rightarrow \pi ^{+}\pi ^{0}n$ more
significantly than the $\gamma p\rightarrow \pi ^{+}\pi ^{-}p$. Then the
cross sections in other isospin channels were calculated using the fixed
parameters. The parameters determined in this way are given in Table 1.

With
this parameter set, the total cross sections of $\gamma N\rightarrow \pi
\pi
N$ can be almost reproduced except the high energy region of the $\gamma
p\rightarrow \pi ^{+}\pi ^{0}n$ and $\gamma n\rightarrow \pi ^{-}\pi
^{0}p$
cross sections and the magnitude of the $\gamma p\rightarrow \pi ^{0}\pi
^{0}p$ cross section. The calculated cross sections (dashed lines) of 
$\gamma p\rightarrow \pi ^{+}\pi ^{-}p$, 
$\gamma p\rightarrow \pi ^{+}\pi ^{0}n$ 
and $\gamma n\rightarrow \pi ^{-}\pi ^{0}p$
are shown in Fig. 2.

The signs of the coupling constants such as $f_{\pi \Delta
N^{*}}^{s}>0$, 
$f_{\pi\Delta N^{*}}^{d}<0$ and $f_{\rho NN^{*}}<0$ were necessary to
explain 
the energy dependence of the $\gamma p\rightarrow \pi ^{+}\pi ^{-}p$
cross section.
The sign of 
$f_{\pi \Delta N^{*}}^{s}$ should be positive to get the constructive
interference between $T_{\Delta KR}$ and $T_{N^{*}\pi \Delta }^{s}$
below
the resonance energy of $N^{*}(1520)$\cite{oset}.
The negative sign of $f_{\rho NN^{*}}$
was also necessary to simultaneously reproduce the cross sections for all
isospin channels of $\gamma N\rightarrow \pi \pi N$ reactions. As far as
the
total cross sections are concerned, the results of our model have been
almost satisfactory except the double neutral pion photoproduction.

The recent experiment on the invariant mass spectra of the $\gamma
n\rightarrow \pi ^{-}\pi ^{0}p$ reaction\cite{exp4} provides 
an opportunity to test
the validity of our model and whether the determined parameters are
appropriate or not.
Although the data have not been published yet since they are still
preliminary,
there seem to be two interesting features from a qualitative
point of view: the first one is the strong correlation at the larger
invariant mass region in the invariant mass spectra of the ($\pi \pi $)
system and the second one is the strong signal of the $\Delta $
resonance in
the invariant mass spectra of the ($\pi N$) system.
We have calculated
the
invariant mass spectra with the parameter-set obtained previously and 
the results (dotted lines) at 730 MeV are shown in Figs. 3(a)-3(b). 
In the ($\pi \pi $) invariant mass spectrum [see Fig. 3(a)], we find
that 
there are a strong peak and small bump coming from 
$\rho N$ production compared with a uniform phase space distribution
(thin solid line).

However, there are no such two peaks in the
experimental
spectrum\cite{exp4}. The theoretical result does not qualitatively agree 
with the experiment. In this calculation, the bump at the
small invariant mass is due to the $\rho$ Kroll-Ruderman term and the
peak at the large invariant mass is related to the $N^{*}$
production
following the decay to the $\rho N$ system.
To further investigate
the origin of this discrepancy, we have also calculated the invariant
mass
spectrum of the ($\pi \pi $) system for the $\gamma p\rightarrow \pi
^{+}\pi
^{-}p$ reaction at 750 MeV[see the dotted line in Fig. 3(c)]. 
Again, we find a strong
peak at the large invariant mass which was not observed in the
experimental spectrum\cite{exp5}. 
This peak is caused by the constructive interference
between the $T_{\Delta KR}$ and $T_{N^{*}\rho N}$ terms. We note that
the $\rho$ Kroll-Ruderman term does not contribute to 
the $\gamma p\rightarrow \pi ^{+}\pi ^{-}p$ reaction.
 These discrepancies indicate that the $\rho NN^{*}$
coupling constant $f_{\rho NN^{*}}$ should be changed from the negative value
to the positive one and the 
$\rho$ Kroll-Ruderman contribution is too large at the small invariant
mass region.

Taking these results into consideration, we modify the parameters to
reproduce the total cross sections of the $\gamma p\rightarrow \pi
^{+}\pi
^{-}p$ and $\gamma p\rightarrow \pi ^{+}\pi ^{0}n$ reactions. The
parameter-set obtained is shown in Table 1, where the sign of 
$f_{\rho NN^{*}}$ is positive and the range parameter $q_{\rho \pi \pi
}$ 
is taken to be larger than the previous one. With these
parameters, the cross sections and the invariant mass spectra 
are calculated
and the results are shown in Figs. 2 and 3, respectively. 
The full calculations (thick solid lines) are consistent with the data
except
the magnitude of the $\gamma n\rightarrow \pi ^{-}\pi ^{0}p$ cross
sections in the higher energy region. The contributions of the $\pi
\Delta$
channel(dashed lines), the $\rho N$ channel arising from the $N^{*}$
production (long dashed lines) and the $\rho$ Kroll-Ruderman term 
(dash dotted lines) are
also plotted in Fig.3, respectively. 
The calculated invariant mass spectra at the other photon energies seem
to
almost consistent with the data from a qualitative point of view.
In the ($\pi \pi$) invariant mass spectrum for $\gamma
n\rightarrow \pi ^{-}\pi ^{0}p$, one can observe that 
the peak shifts to 
the larger invariant mass compared with that of the uniform phase space 
distribution[see Fig.3(a)]. This is due to the $N^{*}$ production
following the decay to the $\rho N$ system. The strong peak at the
$\Delta $
resonance energy in the ($p \pi^{0}$) invariant mass spectrum [see
Fig. 3(b)]
can be seen in the
calculation and is attributed to the strong transition into the $\pi
\Delta $
state. These features in our theoretical results are also found in the
data\cite{exp5}.
Furthermore, the $\rho$ Kroll-Ruderman term is necessary to reproduce the
magnitude of the cross sections for the neutral pion production such as
the
$\gamma p\rightarrow \pi ^{+}\pi ^{0}n$ and $\gamma n\rightarrow \pi
^{-}\pi
^{0}p$ reactions. In the calculations with the modified parameter-set,
the
diagram of
Fig. 1(f) is also included. This diagram contributes to the $\gamma
p\rightarrow
\pi ^{0}\pi ^{0}p$ cross section significantly and leads to the
improvement
of the calculation. We note that the $\rho N$ intermediate state 
does not contribute
to the double neutral pion photoproduction. Since the discussion 
regarding this isospin channel is 
beyond the scope of this letter, we will report the results elsewhere.
For
the other isospin channels, this diagram modifies the cross sections
slightly.

Finally, we discuss the disagreement between the calculation and
the
data in the $\gamma n\rightarrow \pi ^{-}\pi ^{0}p$ cross section. In
this experiment,
the cross sections have been measured by using the detector
with a smaller acceptance ($\leq$ 50\%) compared with other isospin
channels\cite{exp2}. 
The total cross sections have been obtained by extrapolating the
data using either a uniform phase space distribution or the Murphy-Laget
model\cite{laget}.
Since the experimental invariant mass spectra are deviated from the
pure phase space distribution and
the latter model underestimates the magnitude of the cross sections,
the extrapolation procedure may be questionable.
To demonstrate this ambiguity, we calculate the cross section
integrated over the acceptance of the detector and extrapolate it by
using the uniform phase space distribution to get the total cross
section.
As is seen
in Fig. 2(c), our extrapolated cross sections (dash dotted line) are 
in good agreement with the data.

We find that our model with the modified parameters can explain the
total
cross sections of the two-pion photoproduction except the $\gamma
n\rightarrow \pi ^{-}\pi ^{0}p$ cross sections fairly well and reproduce
the
characteristic behavior as mentioned above in the invariant mass
spectra.
These results confirm our previous findings that the $\rho N$ channel
plays
an important role in the two-pion photoproduction as well as the $\pi
\Delta $ channel, especially in the $\gamma p\rightarrow \pi ^{+}\pi
^{0}n$
and $\gamma n\rightarrow \pi ^{-}\pi ^{0}p$ reactions.  The experimental
cross sections of the $\gamma n\rightarrow \pi ^{-}\pi ^{0}p$ reaction 
obtained by the extrapolation procedure are
largely model-dependent as is inferred in the above discussion. In order
to
examine a theoretical model, one should compare the theory directly with
the
data integrated over the acceptance of the detector. 

\acknowledgements
We thank Prof. P. Pedroni for providing us with the data collected at
Mainz
prior to publication.

%
%
\newpage
\begin{table}[H]
\caption{
The parameters used in our model. 
The old parameter-set corresponds to the parameter-set(II) in Ref.
\protect\cite{ochi}. The definitions of the parameters are
described in the text and Ref.\protect\cite{ochi}. 
($^{a}$This value is the sum of the vector and tensor coupling 
constants.)
}
\label{table:coupling}
\begin{center}
\begin{tabular}{|c||c||c|}
 & Old parameter-set & New parameter-set
\\ 
\hline
$ M_{N^{*}}${\footnotesize(MeV)} & 1554 & 1566 
\\
\hline
$f_{\pi N N^{*}} $         &  1.13 & 1.13
\\ 
$p_{\pi N N^{*}}$ {\footnotesize(MeV/c)}       &   400 & 400
\\ 
\hline
$f_{\pi \Delta N^{*}}^{s}$ &  $0.992$ & $0.992$
\\ 
$p_{\pi \Delta N^{*}}^{s} $ {\footnotesize(MeV/c)}      &   200 & 200
\\ 
\hline
$f_{\pi \Delta N^{*}}^{d}$ &  $-1.00$ & $-1.00$
\\ 
$p_{\pi \Delta N^{*}}^{d}$ {\footnotesize(MeV/c)}       &  300 & 300
\\ 
\hline
$f_{\rho N  N^{*}}$                 &  $-0.928$ & $0.583$
\\ 
$p_{\rho N N^{*}}$ {\footnotesize(MeV/c)}       &  200   & 300
\\ 
\hline
$f_{\rho \pi  \pi}$      & 82.0 & 25.6
\\ 
$q_{\rho \pi \pi}$ {\footnotesize(MeV/c)}       & 100   & 200
\\ 
\hline
{\footnotesize $Q_{\Delta}(N \rightarrow \pi \Delta)$ (MeV/c)}& 400 &480
\\ 
{\footnotesize $Q_{\Delta}(\Delta \rightarrow \pi N)$ (MeV/c)}& 358 &358
\\ 
\hline
$G_{T}$ & 17.6 & 21.05$^{a}$
\end{tabular}
\end{center}
\end{table}
\newpage
\begin{figure}[h]
\caption{Diagrams for the two-pion production.
(a) The $\Delta$ Kroll-Ruderman term.
(b) The $\Delta$ pion-pole term.
(c) The $N^{*} \rightarrow \pi \Delta$ contribution.
(d) The $N^{*} \rightarrow \rho N$ contribution.
(e) The $\rho$ Kroll-Ruderman term.
(f) The $\pi \Delta$ production accompanied by the nucleon exchange.
}
\label{fig:fig1}
\end{figure}
\begin{figure}[h]
\caption{The total cross sections of 
(a) the $\gamma p\rightarrow \pi ^{+}\pi ^{-}p$,
(b) $\gamma p\rightarrow \pi ^{+}\pi ^{0}n$
and (c) $\gamma n\rightarrow \pi ^{-}\pi ^{0}p$ reactions.
The solid lines are the calculations with a new parameter-set. 
The dashed lines are the calculations with an old parameter-set.
The dash dotted line in the $\gamma n\rightarrow \pi ^{-}\pi ^{0}p$
reaction
is the extrapolated cross section(see the text for detail).
Experimental data are taken from
Refs.\protect\cite{exp1,exp2,exp5,exp6,exp7}
}
\label{fig:fig2}
\end{figure}
\begin{figure}[H]
\caption{The invariant mass spectra of  
(a) the ($\pi^{-} \pi^{0}$) and (b) ($p \pi^{0}$) systems for
the $\gamma n\rightarrow \pi ^{-}\pi ^{0}p$ reaction at 730 MeV and 
(c) the ($\pi^{+} \pi^{-}$) system for
the $\gamma p\rightarrow \pi ^{+}\pi ^{-}p$ reaction at 750MeV,
respectively.
The thick solid lines are the full calculations with a new
parameter-set.
The contributions of the $\pi \Delta$
channel(dashed lines), the $\rho N$ channel arising from the $N^{*}$
production (long dashed lines) and the $\rho$ Kroll-Ruderman term 
(dash dotted lines) are also plotted. The dotted lines are the full
calculations with an old parameter-set. The calculated cross sections
are obtained by integrating over all phase space.
Experimental data\protect\cite{exp5} are appropriately normalized.
}
\label{fig:fig3}
\end{figure}
\newpage
\pagestyle{empty}
\begin{figure}[H]
\begin{center}
\begin{tabular}[h]{cc}
\leavevmode\psfig{file=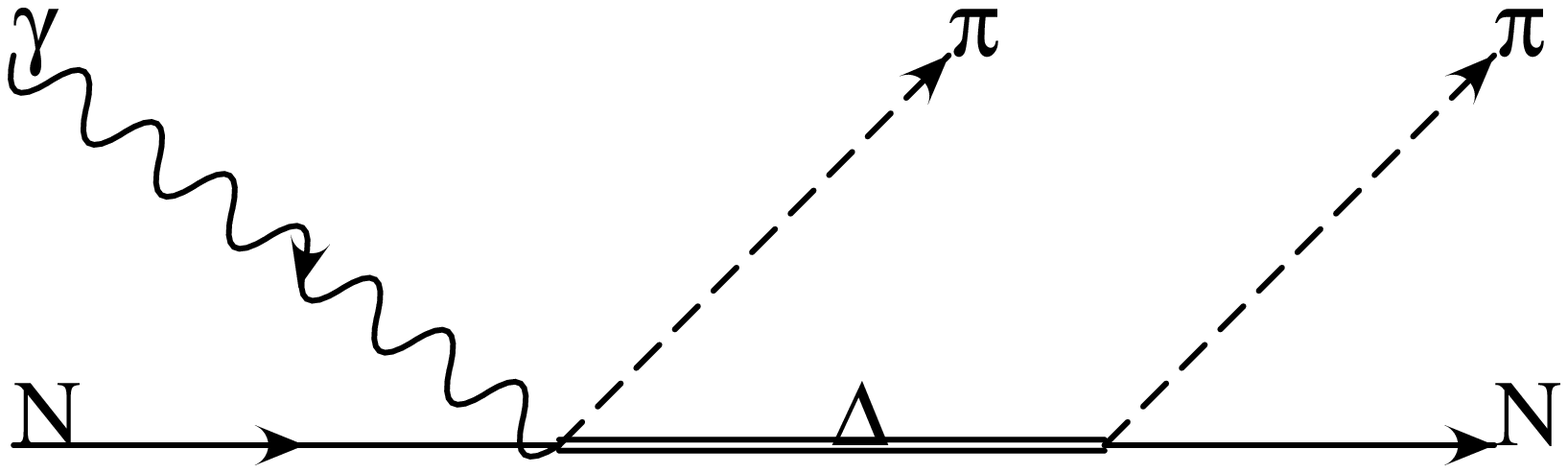,width=6cm}
&
\leavevmode\psfig{file=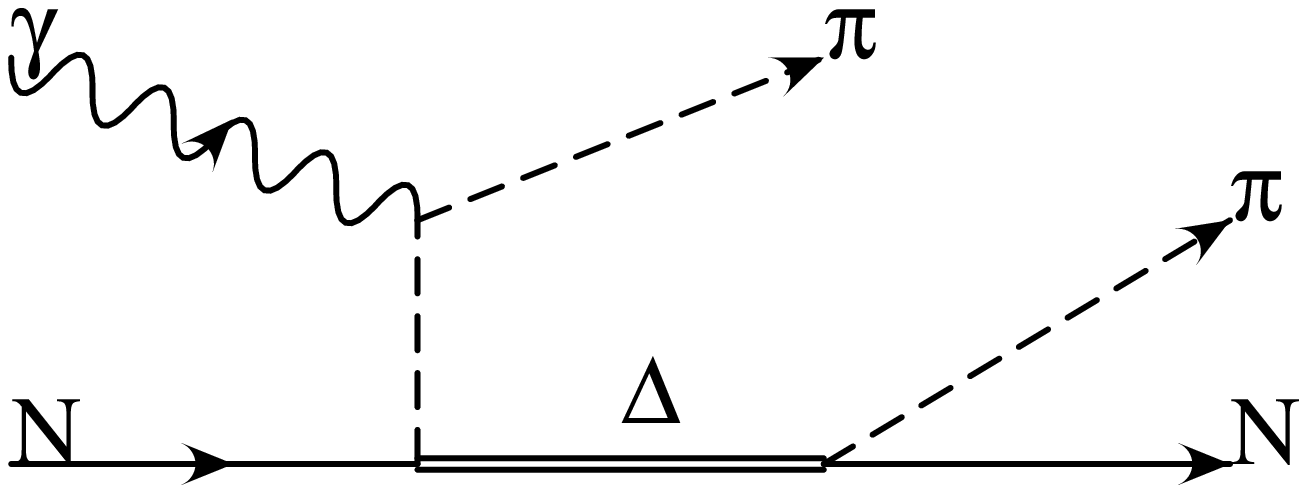,width=6cm}
\\
(a) & (b)
\\
\leavevmode\psfig{file=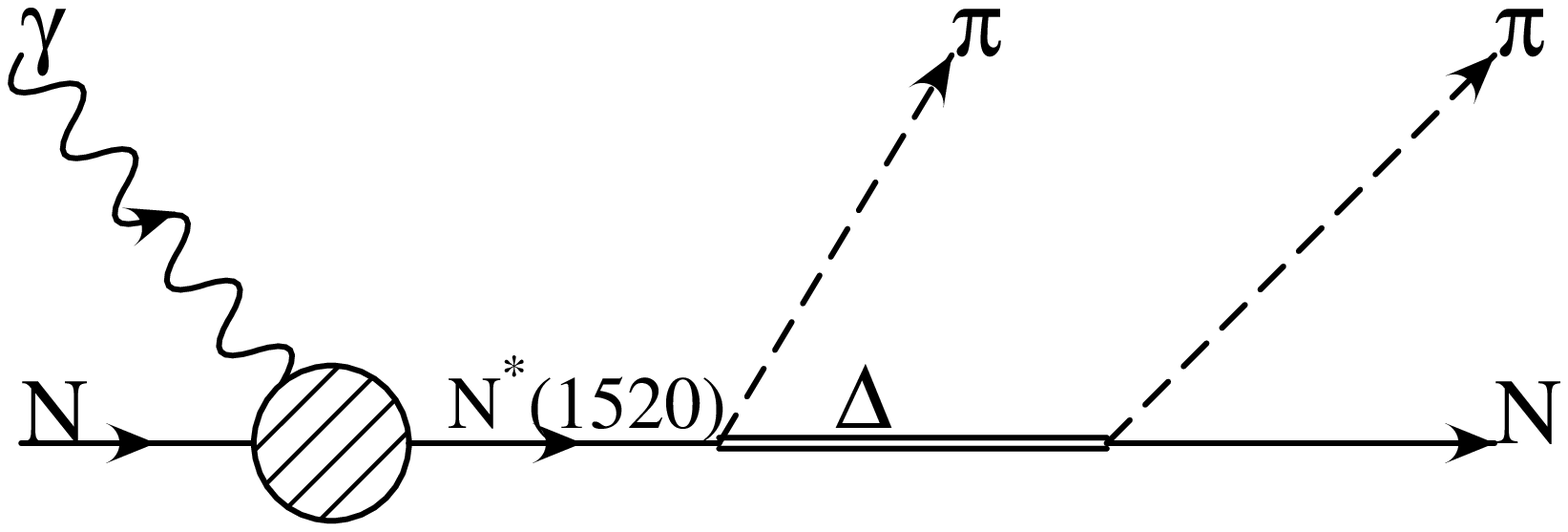,width=6cm}
&
\leavevmode\psfig{file=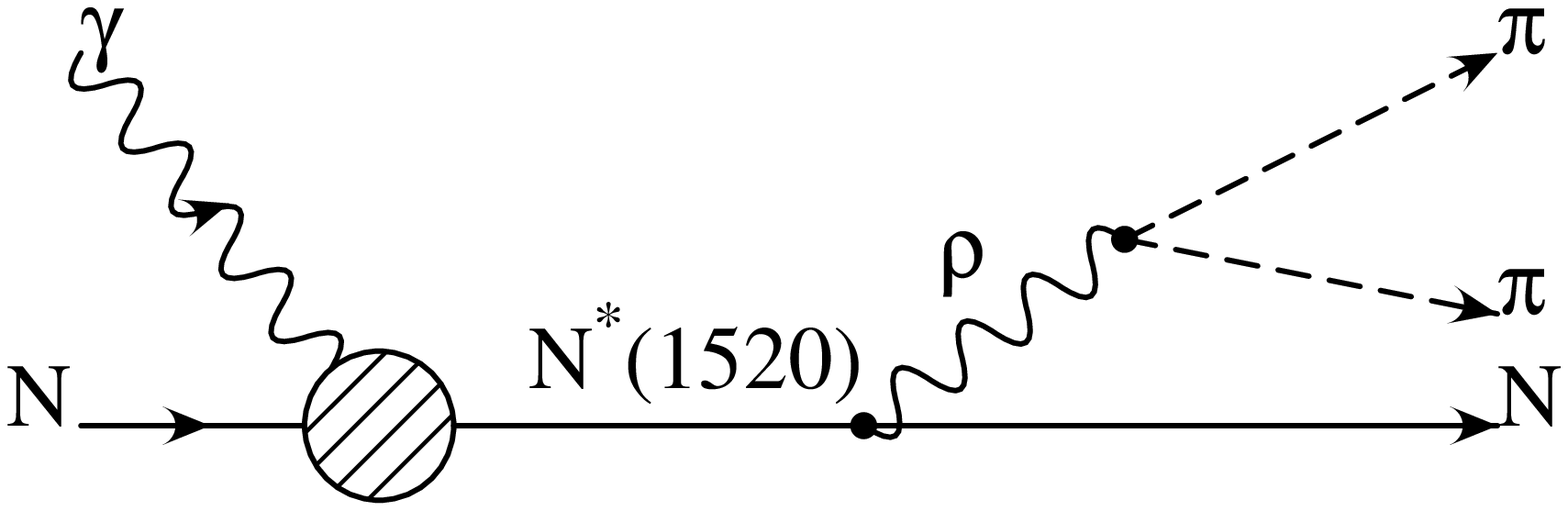,width=6cm}
\\
(c) & (d)
\\
\leavevmode\psfig{file=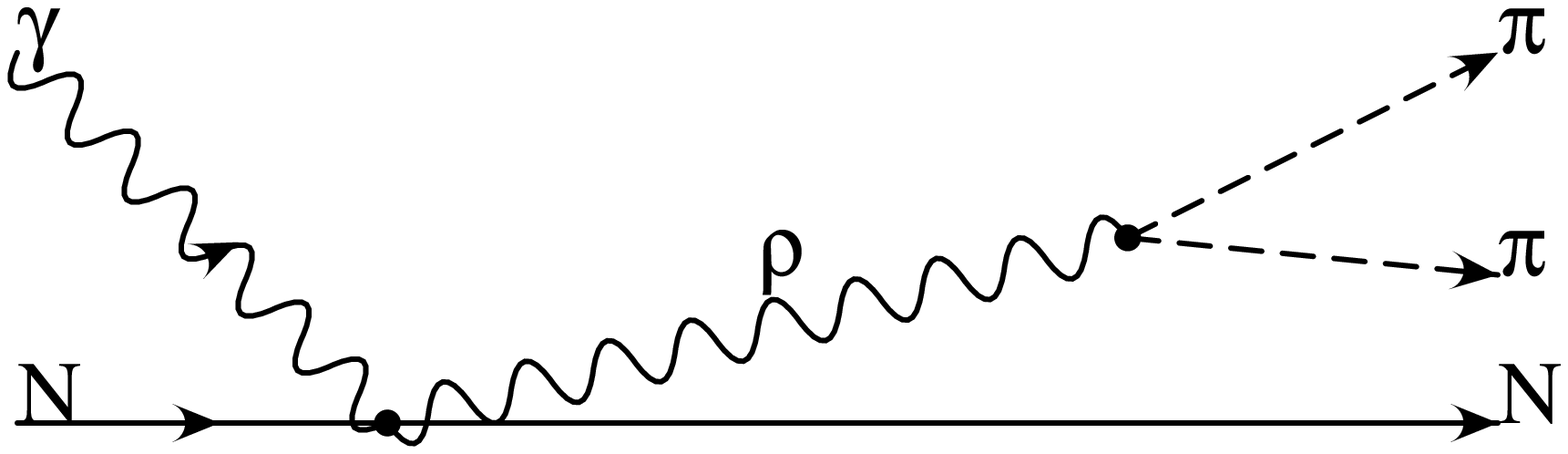,width=6cm}
&
\leavevmode\psfig{file=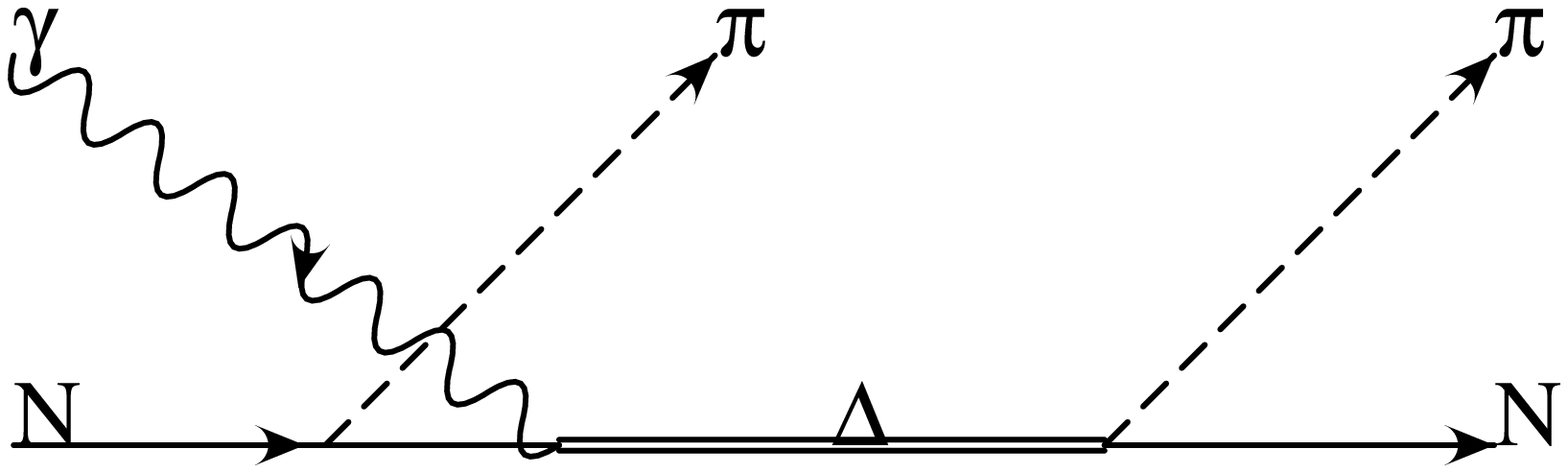,width=6cm}
\\
(e) & (f)
\end{tabular}
\end{center}
\begin{center}
Figure \ref{fig:fig1}
\end{center}
\end{figure}
\newpage
\begin{figure}[H]
\leavevmode\psfig{file=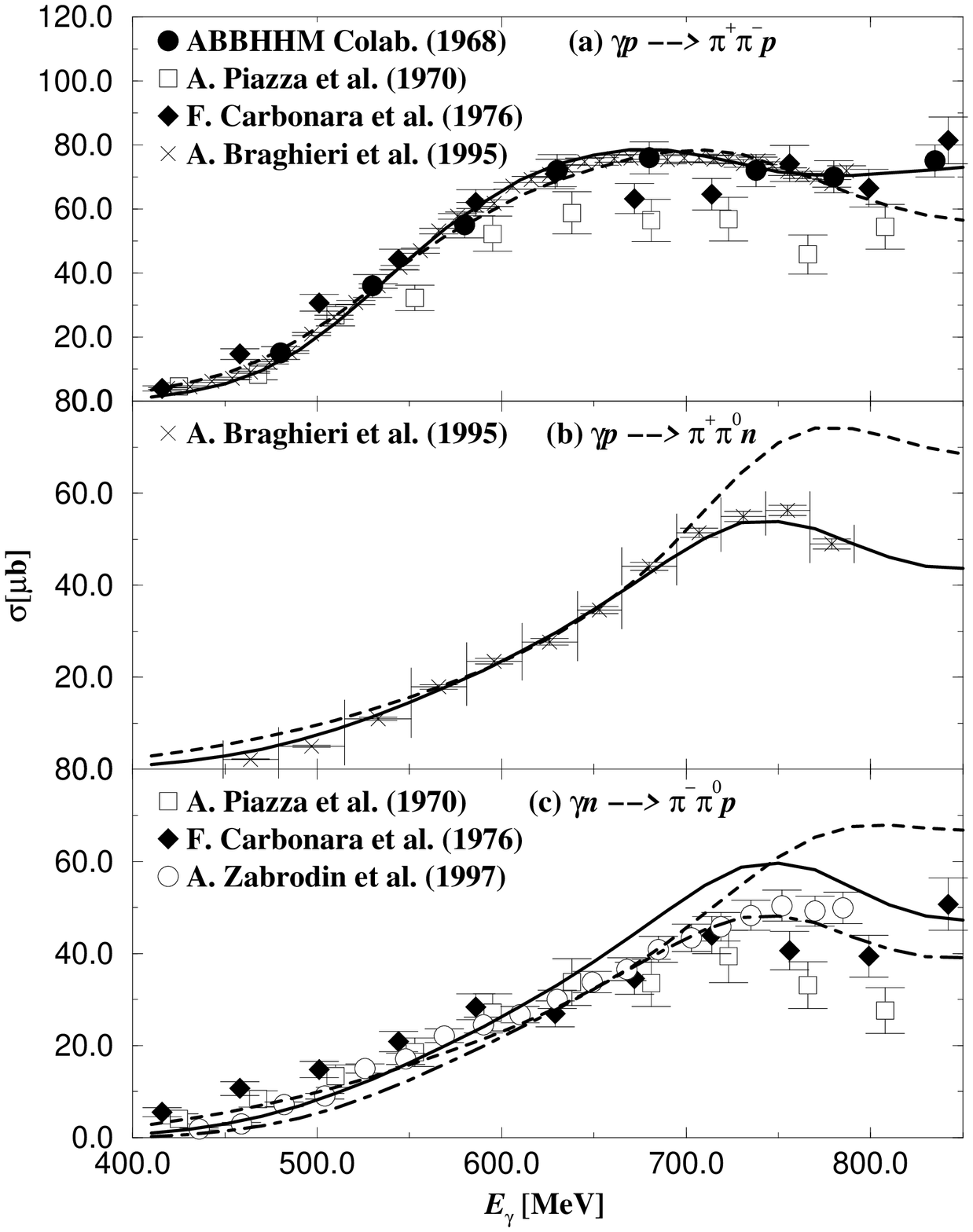,height=22cm,width=17cm}
\begin{center}
Figure \ref{fig:fig2}
\end{center}
\end{figure}
\newpage
\begin{figure}[H]
\leavevmode\psfig{file=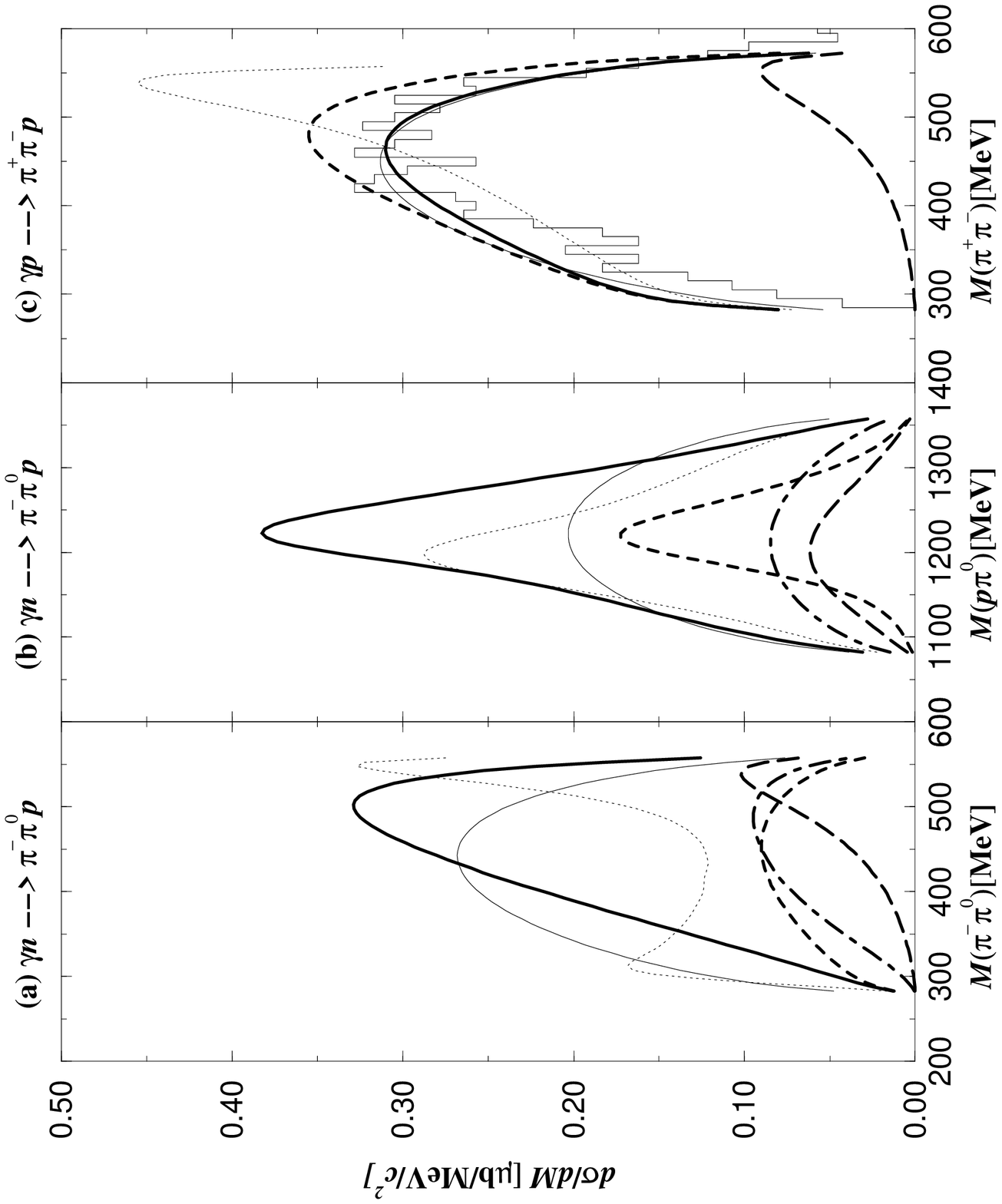,height=22cm,width=17cm}
\begin{center}
Figure \ref{fig:fig3}
\end{center}
\end{figure}
\end{document}